\documentclass[aps,pre,reprint,showpacs,superscriptaddress
]{revtex4-1} 
\usepackage{graphicx}
\usepackage{amsfonts}
\usepackage{amsmath}
\usepackage{amssymb}
\usepackage{txfonts} 
\usepackage{times}
\usepackage{subfigure}
\usepackage{textcomp}
\usepackage{color}
\definecolor{gray}{gray}{0.8}
\usepackage{dcolumn}
\usepackage{bm}
\usepackage{hyperref}

\newcommand{\beq}{\begin{equation}}     \newcommand{\eeq}{\end{equation}}
\newcommand{\beqa}{\begin{eqnarray}}    \newcommand{\eeqa}{\end{eqnarray}}
\newcommand{\bde}{\begin{description}}  \newcommand{\ede}{\end{description}}
\newcommand{\ben}{\begin{enumerate}}    \newcommand{\een}{\end{enumerate}}
             
\newcommand{\noi}{\noindent\mbox{}}

\newcommand{\eqn}[1]{{\beq {#1} \eeq}}

\newcommand{\inv}[1]{{\frac{1}{#1}}}

\newcommand{\inRbracket}[1]{{\left({#1}\right)}}
\newcommand{\inSbracket}[1]{{\left[{#1}\right]}}

\newcommand{\modif}[2]{#2}

\begin{document}
\title{ A mechanism of macroscopic rigid-body behavior through evanescent mode}

\author{ Ken Sekimoto}\affiliation{Mati\`{e}res et Syst\`{e}mes Complexes, CNRS-UMR7057, Universit\'e de Paris, 75013 Paris, France}  \affiliation{ Gulliver UMR CNRS 7083, ESPCI Paris, Universit\'e PSL, 75005 Paris, France}  \author{Yanis Mehdi Benane}\affiliation{{Licence 3, UFR Physique, Universit\'e de Paris, 75013 Paris, France}}
 \author{Karim Elayoubi Alloubia}\affiliation{{Licence 3, UFR Physique, Universit\'e de Paris, 75013 Paris, France}} \author{Romain Arteil}\affiliation{{Licence 3, UFR Physique, Universit\'e de Paris, 75013 Paris, France}} \author{Antoine Fruleux}\affiliation{RDP, Universit\'e de Lyon, ENS de Lyon, UCB Lyon 1, INRAE, CNRS, 69364 Lyon Cedex 07, France } \affiliation{LadHyX, CNRS, Ecole polytechnique, Institut Polytechnique de Paris, 91128 Palaiseau Cedex, France}

\begin{abstract}
 It has been numerically found that the setup of Newton's cradle can exhibit a rigid-body like behavior,
 that is, the target cluster starts to move collectively without emitting the outermost particle if the purely repulsive interaction between the neighbouring particles is very ``soft'' [KS,  Phys. Rev. Lett. 104, 124302 (2010)]. 
We show theoretically that such an 
interaction leads to an evanescent mode within the cluster, which delivers the momentum within the cluster \modif{[6]}{without any propagating oscillation.}
\end{abstract}
\pacs{
{
45.50.-j, 
05.45.-a, 
03.65.Nk, 
45.50.Tn, 
01.50.Wg	  
}
} 

\maketitle
\section{ Introduction} 
Rigid-body behavior is a collective movement of spatially distributed masses.
A common idea of modelling the rigid-body will be a group of masses joined together by extremely hard springs. 
In the limit of large stiffness of the springs, the elastic waves will rapidly propagate to-and-fro being reflected by the boundaries of such mass-and-spring network, resulting in the proper distribution of translational momenta among the masses. 
In the present paper, we will present the other possible mechanism of rigid-body behavior in the context of macroscopic and one-dimensional dynamics, the mechanism in which non-propagating evanescent modes play the central role.
\begin{figure}[b]
\centering
\subfigure[ \null] 
{   \label{fig:schema}\hspace{0.5cm} \includegraphics[width=1.2cm,angle=-90.]{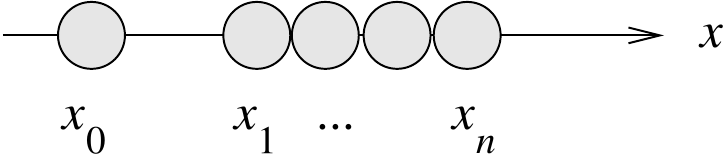}}
\hspace{-0.5cm}
\subfigure[ \null] 
{   \label{fig:hard1sur4}
  \includegraphics[width=4.2cm]{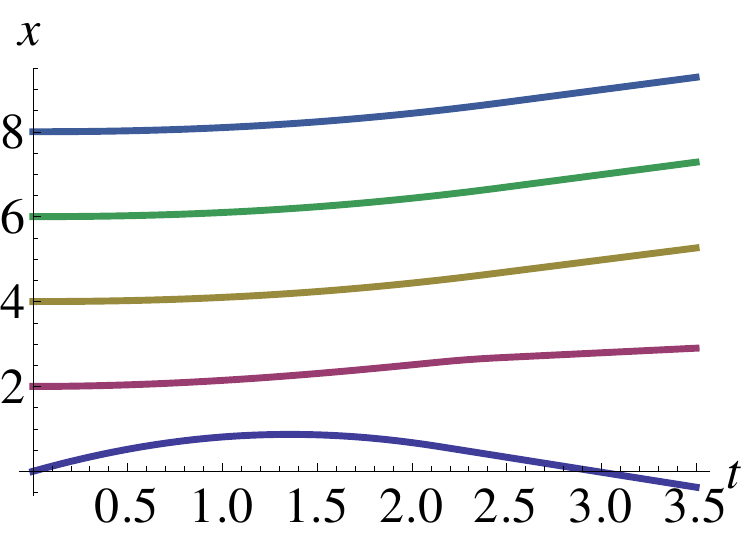}}
  \hspace{3.5cm}
\subfigure[ \null] 
{   \label{fig:1c}
  \includegraphics[width=4.2cm]{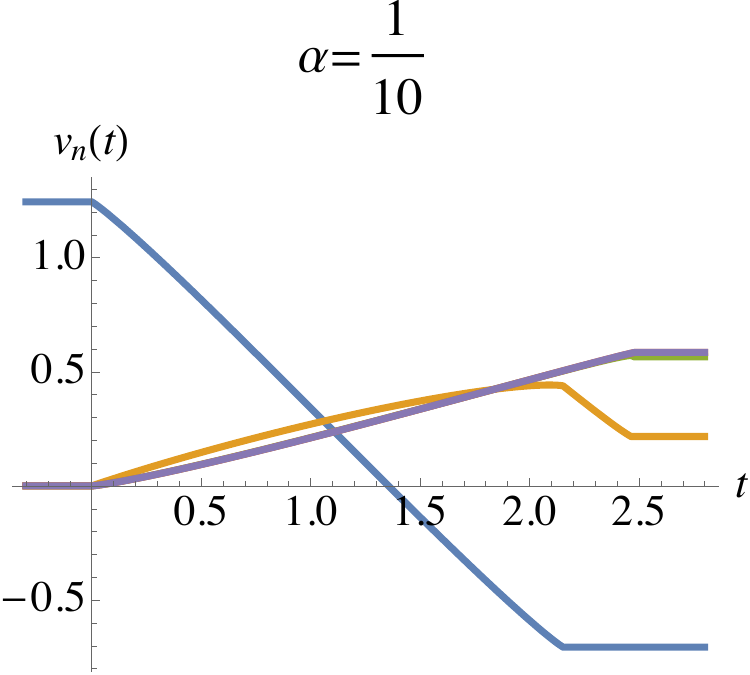}}
  \hspace{-0.2cm}
\subfigure[ \null] 
{   \label{fig:1d}
  \includegraphics[width=4.2cm]{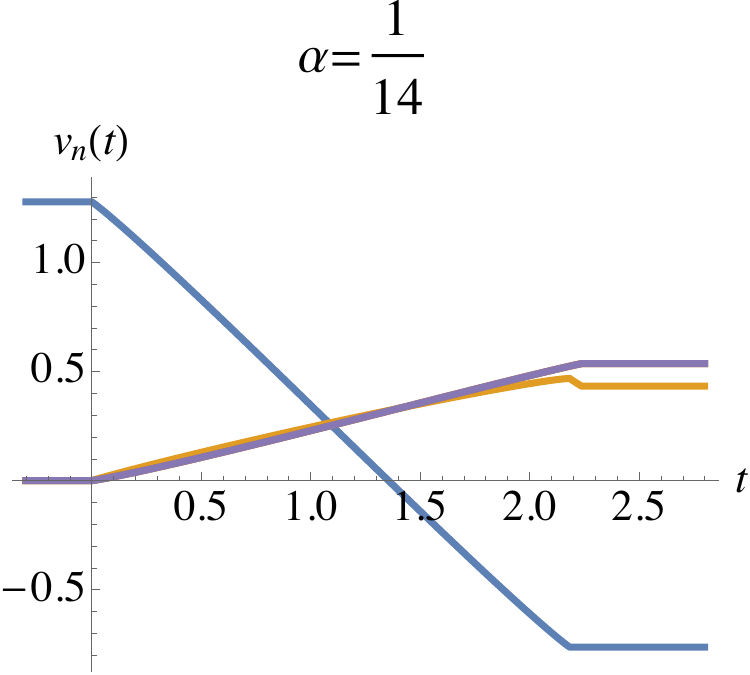}}
\caption{(Color on line)
{\it (a)} An example of numerical setup and result: A hitting particle (position $x_0$) comes into contact 
at $t=0$ with the target cluster of particles (positions, $x_1,\ldots,x_n$, here $n=4$), which have been just in touch for $t<0.$ (cf. With suitable scaling of space and time the incoming velocity can be normalized.)
{\it (b)} 
Results of direct simulation of one-on-four collision ($n=4$) \cite{wagon-prl2010}. The parameters chosen (see (\ref{eq:force-disp}) in the text) are  $m_i=f_{i,i+1}^{(0)}=1$ and $\alpha_{i,i+1}=1/10$ for all $i.$
The height of each curve {shows} the position of a particle,  $x_0,x_1,\ldots,x_4$ from bottom to top, as function of time, $t,$ the horizontal coordinate.
\modif{[7]}{{\it (c)} and {\it (d)}: The velocity of the hitting particle $v_0(t)$ (blue) and those in the target cluster
vs time for $\alpha=\inv{10}$ (c) and for $\alpha=\inv{14}$ (d). The velocity of the front target particle,
$v_1(t)$ (orange), shows {well-visible deviations} from the remaining target particles. {However,} this deviation decreases with the decrease in the exponent $\alpha,$ and should eventually vanish in the limit of $\alpha\to +0,$ according to Fig.3 of \cite{wagon-prl2010} for $n=2$ case.}
}
\label{fig:prl} 
\end{figure}
A decade ago, we discovered numerically a macroscopic rigid-body behavior under {a} setup similar to the Newton's cradle that contains no attractive or cohesive interactions, see Fig.~\ref{fig:schema}  \cite{wagon-prl2010}. 
If the chain of masses interacts only through short-ranged (contact) repulsive forces, we would expect that the mass at the furthest end will detach when the propagating wave brings to it an outward momentum. In other words{,} the elastic waves require attractive forces in order to be reflected at the free boundaries.
The particularity of our system, however, was that the repulsive force obeys {a} power-law $\propto \delta^\alpha$ of the `overlap of the displacements,' $\delta,$ with a very small positive exponent $\alpha$ \cite{wagon-prl2010}. With such {a} repulsive interaction{,} the train of masses reacts as if they were a single rigid-body upon being hit by another mass. Fig.\ref{fig:hard1sur4} recapitulates a result of  one-on-four
collision \cite{wagon-prl2010}. {Strictly speaking}, \modif{[7]}{the true rigid-body behavior is attained only in the limit $\alpha\to +0$  and, otherwise, we will use the word ``rigid-body like.''  The particle velocities in the target cluster are not the same, especially after the detachment of the hitting particle,  but those differences decrease {as} $\alpha\to +0${, and}  as shown in Figs.\ref{fig:1c} and \ref{fig:1d} (see also Fig.3 of \cite{wagon-prl2010}).
}
Below, in \S\ref{sec:model}   we define the model system. Then in \S\ref{sec:MC}  we introduce a mode-coupling model specifically devised for our system, where the evanescent mode appears naturally. In \S\ref{sec:mode-coupl-result} we show that the mode-coupling model reproduces quite  well the direct numerical results.
The last section, \S\ref{sec:discussion}, is for the discussion.
In the appendix we {provide} details of calculations (Appendix \ref{app:general}), numerical test of evanescent mode (Appendix \ref{sec:err}) 
 and also {present an} experimental setup in which the train of masses show the rigid-body like motion (Appendix \ref{sec:exp}).

\section{\null{Model}}\label{sec:model}  
As shown in Fig.\ref{fig:schema}{, we index the particles from left to right, labeling their}
positions and masses {as} $\{x_0,x_1,\ldots,x_n\}$ and $\{m_0,m_1,\ldots,m_n\}$, respectively. The displacement of each particle from its initial position is denoted by $u_i(t)=x_i(t)-x_i(0)$ ($i=0,\ldots, n$) and 
the ``overlap'' between the $i$-th and $(i+1)$-th particles is defined 
\beq
\delta_{i,i+1}\equiv   u_i-u_{i+1} 
\eeq
and we adopt the repulsive force, 
\eqn{\label{eq:force-disp}
f_{i,i+1}(\delta_{i,i+1})=
\left\{\begin{array}{ll} 
      f_{i,i+1}^{(0)}{\delta_{i,i+1}}^{\alpha_{i,i+1}}, & \delta_{i,i+1} >0  \\
      0, & \delta_{i,i+1}       \le 0    \\
   \end{array}\right. ,
}
where both $f_{i,i+1}^{(0)}$ and ${\alpha_{i,i+1}}$ are positive. {A} negative ``overlap'' implies the absence of contact.
To make the description concrete, we will focus {on} the case in which a single particle ($i=0$) collides with the cluster of $n$ particles ($i=1,\ldots,n$) as in Fig.\ref{fig:schema}. The generalization to cluster-cluster collision is, nevertheless, straightforward. 
There are $(n+1)$ equations of motion:
\beqa \label{eq:Newton}
m_0 \ddot{u}_0&=& -f_{0,1}(\delta_{0,1}) 
\cr
m_i \ddot{u}_i &=& f_{i-1,i}(\delta_{i-1,i}) -f_{i,i+1}(\delta_{i,i+1}) \qquad (1\le i<n)
\cr
m_n \ddot{u}_n &=& f_{n-1,n}(\delta_{n-1,n}),
\eeqa
with the initial conditions, $u_0(0)=u_1(0)=\ldots =u_n(0)=0$ and $\dot{u}_0(0)>0$ \footnote{As noted in \cite{wagon-prl2010}, the initial velocity of the hitting particle, $v_0=\dot{u}_0(0),$ can be chosen arbitrarily for the uniform system ($m_i=m,$ $f_{i,i+1}=f_{0,1}$ and $\alpha_{i,i+1}=\alpha_{0,1}$) because it {can be} rescaled by changing the unit of time. It also means that there is {\it no linear regime} in the present problem.}.
\modif{[10]}{(The force $f_{i,i+1}$ is the flow rate of momentum and, like the flow of probability or other conserved quantities on the networks, we use the double indices $\{i,i+1\}$ connecting the ``nodes'' (i.e. particles). Concomitantly, we use double index for the associated $\delta$'s and $\alpha$'s. The viewpoint of momentum flow leads naturally to the evanescent mode described below.) }

\section{\null{Mode-coupling reduction}} \label{sec:MC} 
Next we reduce the above equations {of} motion {to} the mode-coupling equations.
While the approach is rather {\it ad hoc} which is valid only {for a} 1D system, the underlying rule is to observe the conservation of momentum flow, as in many other theories to construct hydrodynamic models \cite{MPP}.
We will extract the modes {by} anticipating the rigid-body like motion and check, a posteriori, the consistency of the model.
The first mode is evidently the relative displacement between the hitting particle ($i=0$) and its target cluster. If the latter behaves as a rigid body, 
we may take the position of the particle which is directly hit, i.e., {particle} $i=1$. We introduce, the first mode variable $\Phi$ through 
\eqn{\label{eq:Phi}
\Phi\equiv \delta_{0,1}(=u_0-u_1).
}
We need at least one other mode {to account for} the momentum conservation {\it inside} of the target cluster,  ($\{x_1,\ldots,x_n\}$). 
In order for the rigid-body behavior of this cluster to be observed{,}
we expect that each constituting mass, e.g., $m_i$ receives, to a very good approximation, the momentum at the rate $m_i A${,} where $A$ is the common acceleration of the cluster. Then the force $f_{i,i+1}$ 
at the overlap $\delta_{i,i+1}$ should provide the momentum for all the masses to the right of this overlap, that is, $f_{i,i+1}=(m_{i+1}+\ldots + m_n)A$ with $i=1,\ldots,n-1.$ Eliminating the common $A${,} we find the relation among the momentum flow, or the force, as follows:
\beq \label{eq:QdM-repartition}  
f_{i,i+1}=\frac{m_{i+1}+\ldots+m_n}{m_2+\ldots+m_n}f_{1,2} \quad (i=1,\ldots,n-1),
\eeq
This result suggests that if we take 
\beq 
\psi \equiv \delta_{1,2}\, (=u_1-u_2)
\eeq
as the second mode variable, all the internal overlap variables $\delta_{i,i+1}$
in this mode should be given through Eq.(\ref{eq:QdM-repartition}).
\modif{[12]}{We use the lowercase letter $\psi$ for the second mode to emphasize its literally evanescent nature;
$\max_t |\psi(t)|$ vanishes in the limit $\alpha\to +0$ while the first mode $\Phi(t)$ remains finite  \footnote{But  $\max_t|\psi(t)^\alpha|$ remains finite for $\alpha\to +0$.}.}

For {clarity reasons}, below we analyze the simplest homogeneous case {when} $m_i=f_{i,i+1}^{(0)}=1$ and $\alpha_{i,i+1}=\alpha$ for all $i$ including both the hitting particle ($i=0$) and the target cluster ($i>0$). The treatment of {the} general inhomogeneous case will be described in Appendix.\ref{app:general}.
\modif{[11]}{With a variety of values for the positive parameters $f^{(0)}_{i,i+1}$ and $\alpha_{i,i+1}$ we have numerically verified that the rigid-body like behavior is reproduced as in {the} homogeneous case{,} as long as all the exponents, $\{\alpha_{i,i+1}\},$  are sufficiently small, {and a} precise criterion will be given  below (see 
  Eqs.(\ref{eq:fast-modes0}), (\ref{eq:estimate}) and (\ref{eq:fast-modes}).   }

By substituting the force-overlap relations (\ref{eq:force-disp}) into the asymptotic relationship (\ref{eq:QdM-repartition}), we 
{find for the homogeneous case,}
\eqn{\label{eq:surface-mode0}
\delta_{i,i+1}=
\inRbracket{\frac{n-i }{ n-1 }}^{\inv{\alpha}} 
\psi}
\noi for $\psi >0.$
Substituting (\ref{eq:Phi}) and (\ref{eq:surface-mode0}) into the {Newton's} equations (\ref{eq:Newton}), we have the mode-coupling equations \modif{[1]}{}
\beqa \label{eq:MCeqs0}
\ddot{\Phi}&=&-2 \Phi^\alpha+\psi^\alpha
\cr
\ddot{\psi}
&=&
\Phi^\alpha-\frac{n}{n-1}\psi^\alpha.
\eeqa
\modif{}{The equations of motion for overlaps $\delta_{i,i+1}$ with $2\le i<n$ then become under the mode coupling approximation}
\beq \label{eq:fast-modes0}
\inRbracket{\frac{n-i}{n-1}}^{\inv{\alpha}}\ddot{\psi}=0 \quad (2\le i<n).
\eeq
The initial conditions for the mode-coupling equations (\ref{eq:MCeqs0}) are 
$\Phi(0)=\psi(0)=\dot{\psi}(0)=0$ and $\dot{\Phi}(0)=\dot{u}_0(0)(>0).$\modif{[2]}{}
The last equations (\ref{eq:fast-modes0})\modif{[13]}{, which cannot be imposed as an additional constraint,} 
 serve as consistency criteria for the mode-coupling reduction: We will discuss in more detail in the next section.
The mode $\psi$ deserves the name of {\it evanescent} mode because 
 (\ref{eq:surface-mode0}) implies that the overlaps inside the cluster, $\delta_{i,i+1},$
decays steeply with the ``depth'', $i$. 
\modif{[14]}{In fact the well known inequality, $\ln z\le z-1$ for $z>0,$ implies
$\inRbracket{\frac{n-i}{n-1}}^{\inv{\alpha}}\le \exp\inRbracket{-\frac{i-1}{n\alpha}}.$}
It is by this evanescent mode, instead of propagating elastic waves, that 
the momentum injected by the hitting particle is shared to all the masses in the cluster.

\section{Test of the Mode-coupling model}
\label{sec:mode-coupl-result}
%
\begin{figure}[h!!]
\centering
\subfigure[ \null] %
{   \label{fig:Psi-phi}  \includegraphics[width=4.2cm]{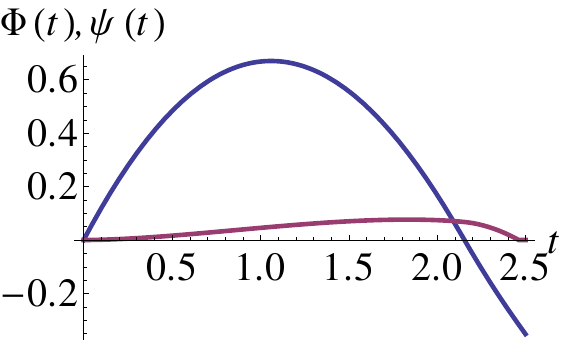}}
\hspace{-0.5cm}
\subfigure[ \null] 
{   \label{fig:force-theo}\hspace{0.5cm}  \includegraphics[width=4.0cm]{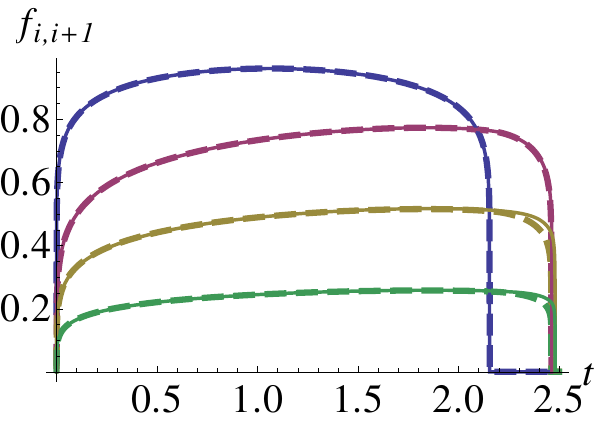}}
%
\subfigure[ \null] 
{   \label{fig:2c}\hspace{0.cm}  \includegraphics[width=4.3cm]{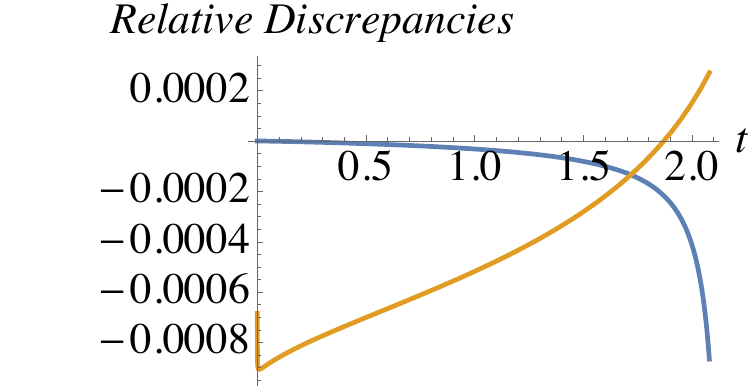}}
\hspace{-0.5cm}
\subfigure[ \null] 
{   \label{fig:2d}\hspace{0.5cm}  \includegraphics[width=3.8cm]{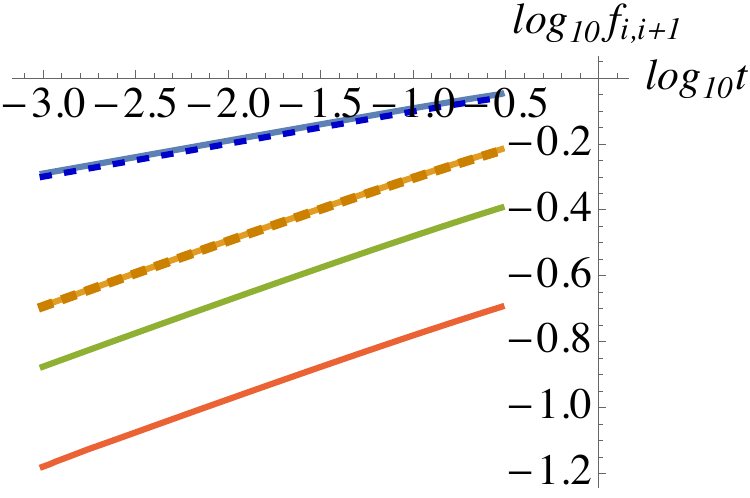}}
\caption{(Color on line)
Results of the mode-coupling model and the comparison with the direct simulation for 
$n=4$ with $m_i=f_{i,i+1}^{(0)}=\dot{u}(0)=1$ and $\alpha_{i,i+1}=1/10$ for all $i.$
{\it (a)} 
Evolution of the center-of-mass mode $\Phi(t)$ (blue, large variation) 
and the intra-cluster {evanescent} mode $\psi(t)$ (red, small variation) obtained by numerically integrating (\ref{eq:MCeqs0}). For $\Phi <0$ the colliding particle $(i=0)$ has detached from the target cluster.
{\it (b)} 
Inter-particle forces, $f_{0,1},f_{1,2},f_{2,3},f_{3,4}$ from top to bottom, as function of time, $t,$ are deduced from the mode-coupling model (\ref{eq:MCeqs0}) and (\ref{eq:surface-mode0}) (thick dashed curves) and are compared with 
the full numerical integration of {Newton's} equations (\ref{eq:Newton}) (thin solid curves).
{\it (c)} \modif{[4]}{Relative discrepancies between the full numerical vs mode coupling
results for $f_{0,1}$ (blue) and $f_{1,2}$ (orange).}
{\it (d)} \modif{[3]}{Log-log plot of the detailed inter-particle forces in the initial regime,
  $f_{0,1}, f_{1,2}, f_{2,3}, f_{3,4}$ from top to bottom. The two dashed curves are the mode-coupling results, which  will be asymptotically
 $f_{0,1} \propto \Phi^\alpha \sim t^{\alpha}$ and $f_{1,2}\propto \psi^\alpha \sim t^{\alpha(2+\alpha)}$ with $\alpha=1/10$ 
 according to (\ref{eq:MCeqs0}).}
}
\label{fig:prl} 
\end{figure}
Fig.\ref{fig:Psi-phi} shows the solution of  (\ref{eq:MCeqs0}) as function of time.
In Fig.\ref{fig:force-theo} we compare  the inter-particle forces $f_{i,i+1}$  ($i=0,1,2,3$)
calculated through Eq.(\ref{eq:surface-mode0}) with  those obtained directly by the full numerical simulation of {Newton's} equations, (\ref{eq:Newton}).	
\modif{[4]}{Fig.\ref{fig:2c} shows, for $f_{0,1}$ (blue) and $f_{1,2}$ (orange), the 
relative discrepancies between the full numerical result $f^{(num)}_{i,i+1}$ and 
the mode-coupling result $f^{(MC)}_{i,i+1},$  that is, $[f^{(num)}_{i,i+1}-f^{(MC)}_{i,i+1}]/f^{(num)}_{i,i+1}$ over the entire time domain of collision. 
The mode-coupling model reproduces the full results up to $10^{-3}$ relative errors}  \modif{[6]}{with no sign of fine oscillations.}
\modif{[3]}{In Fig.\ref{fig:2d} the solid curves show the detailed initial rising of the inter-particle forces, from top to bottom, $f_{0,1}, f_{1,2}, f_{2,3}, f_{3,4}$ by log-log plotting.
We have also shown the mode-coupling predictions for $f_{0,1}$ and $f_{1,2}$ by dashed curves. The {\it absence} of propagative chain waves from the beginning is consistent with the divergence of both linear and soliton wave velocities at vanishing prestress 
\cite{nesterenko-softening-prl2013}.} 
The reproduction by the mode-coupling model is quite satisfactory.
Logically the the hypothesis of the evanescent mode (\ref{eq:surface-mode0}), being equivalent to $f_{i,i+1}={[(n-i)/(n-1)]} f_{1,2}$, can be tested {\it within} the numerical simulation, independently from the mode-coupling equations (\ref{eq:MCeqs0}); we have verified this and the details are given in Appendix \ref{sec:err}.

We also confirmed numerically that the consistency requirement (\ref{eq:fast-modes0}) is satisfied to a good approximation: For example, in the case of  $n=4$ with $m_i=f_{i,i+1}^{(0)}=v_0=1$ and $\alpha_{i,i+1}=1/10$ for all $i,$  the prefactor on the left hand side (lhs) of (\ref{eq:fast-modes0})   is $[(4-i)/(4-1)]^{10}\le 0.02$ for $2\le i\le 4,$ and the numerical
  solution of  (\ref{eq:MCeqs0}) gives $d^2\psi/dt^2 \sim 0.1,$ leading to the lhs of (\ref{eq:fast-modes0}) to be $\lesssim 10^{-3}.$
All these {justify} our choice of the slow variables for the mode-coupling model.

The physical meaning of the lhs of (\ref{eq:fast-modes0}) may deserve to be further clarified:
For simplicity we shall adopt the normalization of $f_{i,i+1}^{(0)}= m_i=1.$
From (\ref{eq:surface-mode0}) \modif{[15]}{the lhs of (\ref{eq:fast-modes0})
reads $\ddot{\delta}_{i,i+1}$ and, therefore, its order of magnitude}
 should be $\sim \delta_{i,i+1}/({\tau_{\rm dur}})^2,$ where ${\tau_{\rm dur}}$ characterises the duration of collision.
On the other hand,  the characteristic \modif{[6]}{response time} $\tau_{i,i+1}$ associated to the overlap $\delta_{i,i+1}$ is such that $(\tau_{i,i+1})\sim {k_{i,i+1}^{\rm (loc)}}^{-\inv{2}},$ where the local stiffness 
at the corresponding overlap, $k_{i,i+1}^{\rm (loc)} \equiv 
 (df_{i,i+1}/d\delta_{i,i+1})$ is estimated to be $\sim {\delta_{i,i+1}}^{-1},$ where we used the approximation $\alpha-1\simeq -1.$  Therefore, the lhs of (\ref{eq:fast-modes0}) reads 
$ \inRbracket{\frac{n-i}{n-1}}^{\inv{\alpha}}\ddot{\psi}\sim  \inRbracket{\frac{\tau_{i,i+1}}{\tau_{\rm dur}}}^2,$ and we can say that the consistency condition (\ref{eq:fast-modes0}) requires the separation of timescales:
\beq \label{eq:estimate}
    \inRbracket{\frac{\tau_{i,i+1}}{\tau_{\rm dur}}}^2 \ll 1.
\eeq %
This equality is evidently fulfilled in the limit of $\alpha\to 0$ as seen in (\ref{eq:fast-modes0}).
In this limit the flows of momentum, $\{f_{i,i+1}\},$ remain finite{,} satisfying (\ref{eq:QdM-repartition}){,} while the amplitude of $\psi(t)$ vanishes \modif{[12]}{as mentioned already}. 
Remark: If we were to take this limit $\alpha\to 0$ from the start, the repulsive force $f_{i,i+1}$ would become undetermined for $0\le f_{i,i+1}\le f_{i,i+1}^{(0)},$ like the Lagrange multipliers, and we should lose sight of the evanescent mode.
Alternatively, it is tempting to regard the target cluster as a kind of mechanical black-box, \modif{[16]}{but the non-potential} structure of Eq. (\ref{eq:MCeqs0}) opposes against naive interpretations in terms {of Newtonian} mechanics.

Apart from the very limit of $\alpha\to 0$, below we identify the condition on $\alpha$ for the validity of the mode-coupling model:
The analysis of (\ref{eq:MCeqs0}) {reveals} that, 
for $\alpha\to 0,$
  (i) the maximum amplitude of $\psi$ {is} decreasing as $\sim e^{-1/(n\alpha)},$ (ii) the duration of contact between the hitting particle and the target cluster decreases as $\tau_{\rm dur} \sim e^{-1/(2n\alpha)}/\sqrt{\alpha}.$ Using these together with the aforementioned estimation, $((n-i)/(n-1))^{1/\alpha}\le e^{-(i-1)/(n\alpha)},$ 
   we find that the lhs of (\ref{eq:fast-modes0}) to be bounded by $\alpha e^{-(i-1)/(n\alpha)} $ for $i\ge 2.$   Thus (\ref{eq:fast-modes0}) holds as a good approximation for  $\alpha \ll n^{-1}.$ 
\null{For $\alpha >n^{-1}$ the rigid-body like behavior does not show up even with $\alpha \ll 1$  \cite{nesterenko-softening-prl2013}.}
In Fig.\ref{fig:Psi-phi} the time at which $\Phi(t)$ returns to zero is advanced with respect to the return of $\psi(t)$ to zero. \modif{[7]}{ Figs.\ref{fig:1c} and \ref{fig:1d} indicates that this discordance will vanish in the limit  $\alpha\to +0.$}

\modif{[5]}{It may be worth distinguishing the rigid body behavior {\it during} the collision from the one {\it after} the detachment of the hitting particle. While the hitting particle injects the momentum through the contact, the relation (\ref{eq:surface-mode0}) with {small}
non-negative $\alpha$   is sufficient to {translate} the target cluster as a (strict) rigid body{,} because (\ref{eq:surface-mode0})  is equivalent to (\ref{eq:QdM-repartition}) for the homogeneous system.
 Once the hitting particle detaches, however, the finite overlaps obeying the very relation (\ref{eq:surface-mode0}) cause the velocity divergence among the cluster particles in non-oscillatory manner, see Figs.\ref{fig:1c} and \ref{fig:1d}.}
\modif{[6]}{It is only in the limit $\alpha\to +0$ that the target cluster does not accumulate the  compressive energies {\it during} the collision, $\sim (\delta_{i,i+1})^{\alpha+1}\to +0,$ (while  the forces  remain finite, $\sim (\delta_{i,i+1})^{\alpha}$), and this limit assures the (strict) rigid body behavior {\it after} the collision.}

\section{Discussion}\label{sec:discussion}

\paragraph*{{R\'esum\'e:}} 
While restrained to a macroscopic and one-dimensional context, 
we have presented a mechanism for the rigid-body like behavior of the particle cluster in which the particles interacts only with truncated repulsive force. 
When the interaction is "soft" enough, the {evanescent} mode distributes the momentum among the constituent particles of the cluster, while the oscillatory modes, \modif{[6]}{which were present 
under finite prestress \cite{nesterenko-softening-prl2013}, are absent.}
The {evanescent} mode also appears in situations other than the collisional setup: If in (\ref{eq:Newton}) we replace the repulsive force by the hitting particle, $f_{0,1}(\delta_{0,1}),$ by an external steady force $f_0 (>0)$ applied from $t=0,$ then a time-periodic {evanescent} mode appears in the cluster on top of the uniform acceleration of the center of mass (data not shown).
 \modif{[9]}{The evanescent mode-coupling is not the unique route to the rigid body as singular limit:
 Different approaches to the rigid body limit show different aspects. 
For a chain of mass and springs, the rigid body limit is attained by the divergent spring constants, where
the oscillatory modes delivers the momentum among the masses.\footnote{The difference between these two approaches is somehow reminiscent of the singular limits of $e^{-\inv{z}}$ in the complex plane; the function is oscillatory when  $z=0$ is approached along the imaginary axis while it is monotonous when approached along the real axis.}}

\paragraph*{{Robustness and limitations:}} 
The rigid-body like behavior is robust against the heterogeneity in the constituent particles as well as the particle-particle interactions, the aspect which is in marked contrast to the ``usual Newton's cradle'' (with $\alpha>1$). In the latter case the propagating soliton modes are highly perturbed by the inhomogeneity \cite{neste-perturbe-Sen-pre1998}. 
The robustness of our system motivated us to try a primitive experimental setup that corresponds approximately to Eq.(\ref{eq:force-disp}) with $\alpha_{i,i+1}=+0,$ where the force $f_{i,i+1}$ tends to a step-function of the overlap $\delta_{i,i+1}.$  See Appendix.\ref{sec:exp}, where we show 
{results from such a} simple experimental set up.

\paragraph*{{Future scope:}} 
Despite the robustness within the one-dimensional models, for the moment we have no idea about possible generalization to more than one-dimensional systems, where 
coupling between linear and angular momenta (cf. \cite{IK-bio-AFKS2016,Valise2017}) {is present}.
The rigid-body behavior without propagating mode is reminiscent of 
the quantum {M\"ossbauer effects} \cite{Mossbauer-original,Kaufman-Lipkin},  the phenomenon of emission or absorption of a photon by a nucleus of metal atom (i.g. of Fe) bound in a crystal lattice {\it without} emission of propagating phonons as recoil at a well characterized probability \cite{Lamb1939}. 
When no phonons are excited, only the center-of-mass of the crystal {exchanges} momentum with the photons, thus the crystal behaves as a rigid body.
In more general context, 
the concept of the  {momentum flow} is sometimes useful like the present case where 
we derived the {evanescent} mode through the analysis of the momentum flow.
 This notion allowed to understand the essential physics of the adiabatic piston \cite{AFRKKS2012} and of the Irving-Kirkwood stress formula, the formula which should be valid even for non-equilibrium and complex systems \cite{IK-bio-AFKS2016}. 
Beyond physics in the narrow sense, the momentum flow may also be a useful notion for practicing sports and playing musical instruments\modif{[17]}{:  In those {areas} the notion {of} ``{making} pass the force through (a part of) the body'' {is often evoked,} and it should be closely related to the momentum flow. The capture of  momentum flow upon these practices will be as informative as the capture of motion.}

\begin{acknowledgments}
We thank Alexandre Di Palma for the technical assistance. KS thanks Motoaki Bamba for his comments. 
\end{acknowledgments}

\appendix
\section{Mode-coupling reduction in the inhomogeneous case}\label{app:general}
We {repeat the steps}
below Eq.(\ref{eq:QdM-repartition}) in the main text except that we  allow here for the heterogeneities in masses $m_i$, force coefficients $f_{i,i+1}^{(0)}$, and exponents $\alpha_{i,i+1}.$

By substituting the force-overlap relations (\ref{eq:force-disp}) into the asymptotic relationship (\ref{eq:QdM-repartition}), we have the representation,  
\eqn{\label{eq:surface-mode}
\delta_{i,i+1}=\inSbracket{\frac{(m_{i+1}+\ldots m_n)f_{1,2}^{(0)}}{(m_{2}+\ldots m_n)f_{i,i+1}^{(0)}}
}^{\inv{\alpha_{i,i+1}}} \psi^{\frac{\alpha_{1,2}}{\alpha_{i,i+1}}}
}
for $\psi >0,$ where 
\beq 
\psi \equiv \delta_{1,2}\, (=u_1-u_2).
\eeq
Eq. (\ref{eq:surface-mode}) implies that the mode $\psi$ is non-propagating and surface localized, that is, {evanescent} mode that decays steeply with the ``depth'', $i$. 
See the main text for the illustration.
The two modes, i.e., the center-of-mass mode, $\Phi,$ and the {evanescent} mode, $\psi,$ will constitute the mode-coupling system.
Substituting (\ref{eq:Phi}) and (\ref{eq:surface-mode}) into the {Newton's} equations (\ref{eq:Newton}), we have
\beqa \label{eq:MCeqs}
\ddot{\Phi}
&=& 
-\inRbracket{\inv{m_0}+\inv{m_1}}\,f_{0,1}(\Phi)+\inv{m_1} {f_{1,2}}(\psi)
\cr
\ddot{\psi}
&=&
\frac{1}{m_1}f_{0,1}(\Phi) -\inRbracket{\inv{m_1}+\inv{m_2+\ldots+m_n}}\,f_{1,2}(\psi)
\eeqa
together with 
\beq \label{eq:fast-modes}
\inSbracket{\frac{(m_{i+1}+\ldots m_n)f_{1,2}^{(0)}}{(m_{2}+\ldots m_n)f_{i,i+1}^{(0)}}}^{\inv{\alpha_{i,i+1}}}
 \frac{d^2}{dt^2}\inRbracket{\psi(t)^{\frac{\alpha_{1,2}}{\alpha_{i,i+1}}}} \, = 0
 \quad (2\le i<n).
\eeq
The logic we adopt is that the mode-coupling equations (\ref{eq:MCeqs}) represent the core of  the rigid-body like motion if (\ref{eq:fast-modes}) holds approximately for small enough exponents, $\alpha_{i,i+1}.$

Some {further generalizations are also conceivable:}
(i) By virtue of {Galilean} invariance the role of the hitting particle and that of the target cluster are exchangeable. 
(ii) The generalization to the cluster-cluster collision is then straightforward; we only have to introduce another {evanescent} mode, say $\overline{\psi},$ for the hitting cluster.
(iii) Because the evanescent mode concerns only the intra-cluster forces,  the repulsive force at the interface between the hitting particle and the target cluster need not satisfy the condition $\alpha\ll n^{-1}.$ 
For example the truncated Hookean spring ($\alpha_{0,1}=1$) works as well (the data not shown). 

\section{Test of the hypothesis of evanescent mode within the numerical simulation \label{sec:err}}
\begin{widetext}
\begin{figure}[h]{}
\centering
{
 \includegraphics[width=5.3 in,angle=-0.]{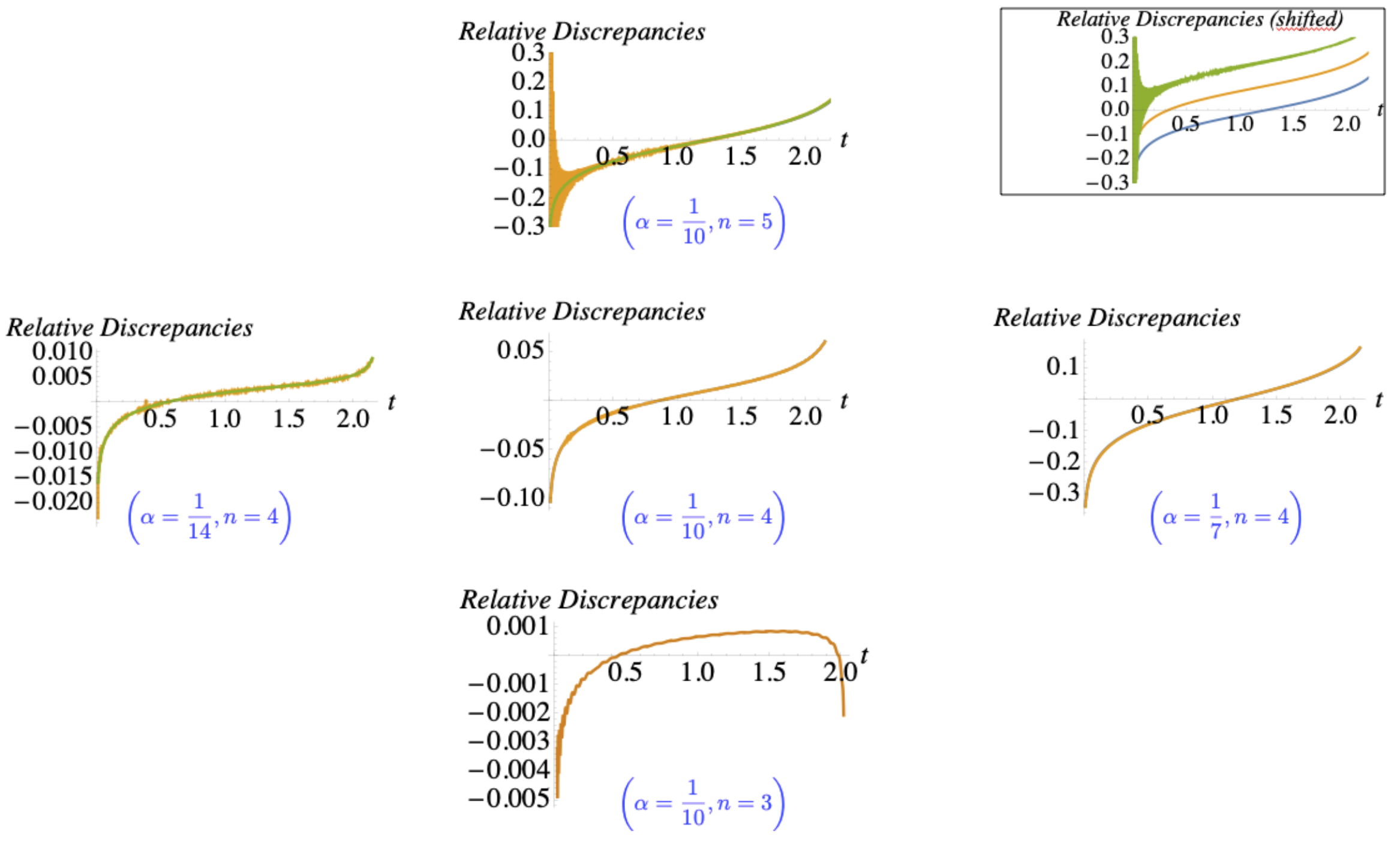}
}
\caption{(Color on line) Test of the evanescent hypothesis (\ref{eq:surface-mode0}) within the numerical simulation. See the main text of Appendix \ref{sec:err} for the description. }
\label{fig:err} 
\end{figure}
 \end{widetext}

 If we recall that $\psi=\delta_{1,2},$ the hypothesis (\ref{eq:surface-mode0}) claims that $\Delta_{i,i+1}=\delta_{1,2}$ holds as a good approximation, where $\Delta_{i,i+1}$ ($i=2,\ldots,n-1$) is defined by $\Delta_{i,i+1}\equiv [(n-1)/(n-i)]^{\inv{\alpha}}\delta_{i,i+1}.$ In Fig.\ref{fig:err} we show the relative discrepancy from this equality, $\Delta_{i,i+1}/\delta_{1,2} -1,$ over the period of collision for various exponents ($\alpha=\inv{7},\inv{10}$ and $\inv{14}$) and also for various $n$, i.e., the number of {particles in the} target cluster ($n=3,4$ and $5$), {and} using only the numerical data.
 Except for the bottom-center figure that shows a single curve
 $\Delta_{2,3}/\delta_{1,2}-1,$  more than one curves coincide within the thickness of the curves. 
 {To demonstrate this, in the inset (top-right) we have vertically offset the  
 two curves 
 $\Delta_{4,5}/\delta_{1,2}-1$ and $\Delta_{3,4}/\delta_{1,2}-1$ with respect to $\Delta_{2,3}/\delta_{1,2}-1$ (from top to bottom)
 shown in the top-center figure $(\alpha=\inv{10},n=5)$.}

 The relative discrepancy changes sign {midway through} the collision and its absolute value increases {at} both ends. 
We should {point out} that the numerical precision is limited especially near the {beginning and end of the duration of the collisions}, which caused a {spurious} 
oscillation for $(\alpha,n)=(1/10,5)$ or a unnatural sign change near the end for $(1/10, 3),$ as well as the noisy data for $(1/14, 4).$
Nevertheless, the overall trend confirms the claim in \S IV (the second last paragraph) that the discrepancy is smaller for the smaller values of $n\alpha,$ especially in the intermediate time range during which the most of the momentum transfer is made. 
  In brief the fairly good applicability of (\ref{eq:surface-mode0}) to our system is observed and the results are consistent with the good reproducibility of the inter-particle forces by the mode-coupling model shown in Fig.\ref{fig:force-theo}.

\section{{Experimental realization of rigid-body behavior}\label{sec:exp}} 
\begin{figure}[b]{}
\centering
{
 \includegraphics[width=3.6 in,angle=-0.]{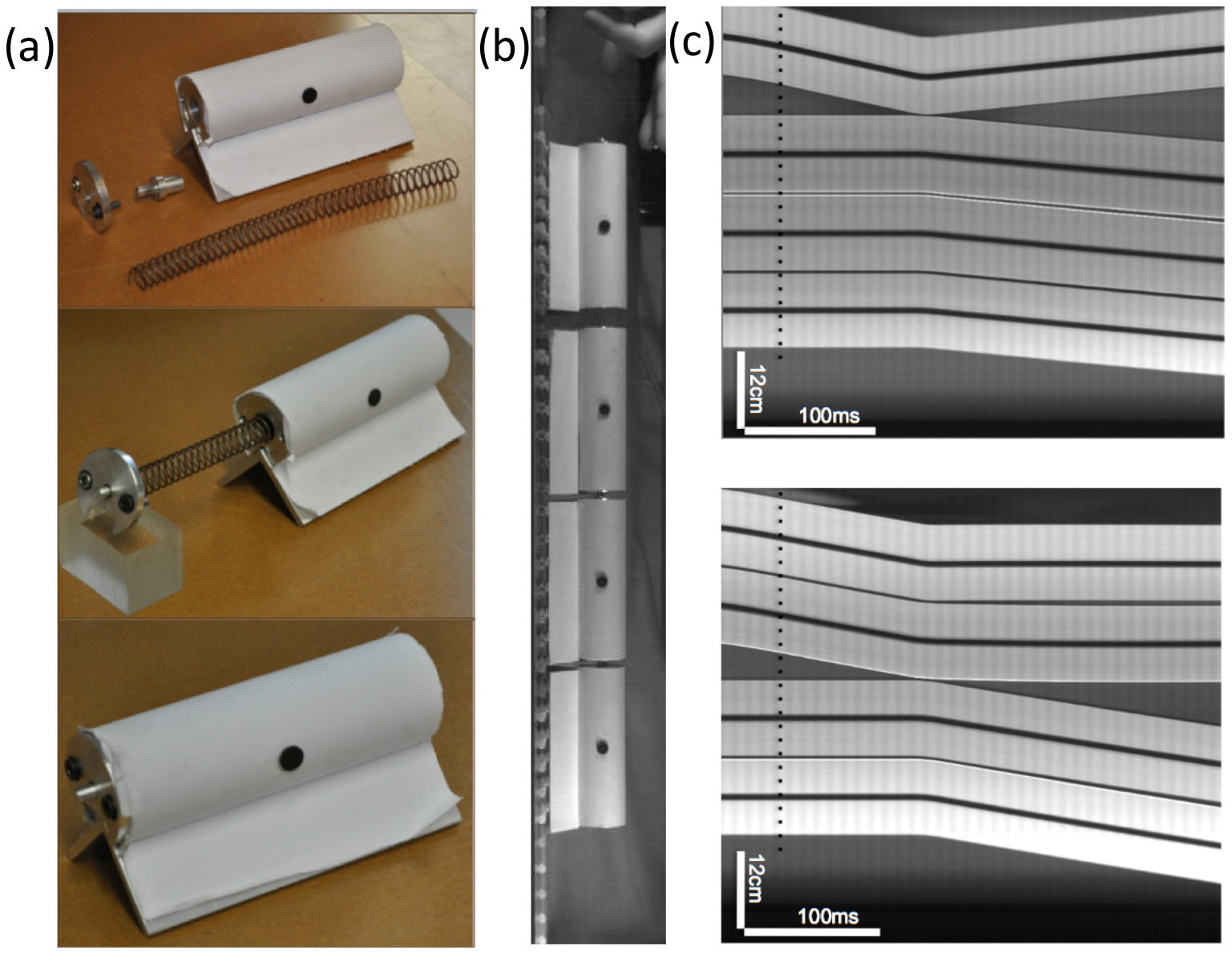}
\hspace{0.2cm}
}
\caption{(Color on line)  Experimental setup and results. 
See the main text for the details.
{\it (a)} Constitution of each ``particle''.
{\it (b)} Arrangement of a ``particle'' colliding against the cluster of three ``particles'' (90$^\circ$ rotated clockwise). 
The outer length of each ``particle'' is   $12.0\pm 0.1{\rm cm}$.
{\it (c)} Kymographical images for the one-on-three collision {\it (Top)}  and the two-on-two collision  {\it (Bottom)}, respectively. The thick curves trace the black markers on the side of ``particles,'' seen in {\it (b)}.
The horizontal and vertical bars on the corner shows the scales of time (100ms) and length (12cm), respectively.
}
\label{fig:data} 
\end{figure}

Fig.\ref{fig:data} (a) shows the architecture of each ``particle'' : 
The main body of each particle is a cylinder, where a pre-compressed spring is confined in the following manner; we attach a piston to one end of the spring, and the piston is constrained by the annular cylinder head. The whole mass of each ``particle'' was within the range of $0.396\pm 0.002$ kg. $\{0.394(5), 0.398(0), 0.396(7),0.393(8)\}$
By this construction, the repulsive force between the neighbouring particles appears only when 
the piston is pushed inwards, which corresponds to $\delta >0.$ The experimental force function, $f(\delta),$ is, nominally,
\eqn{\label{eq:FvsX}
f(\delta)=   \left\{\begin{array}{lr} 
         \mbox{in the range of  $[0,k\Delta_0[$}  & \quad (\delta = 0)    \\
           k(\delta+\Delta_0) &\quad (\delta >0)  \\
   \end{array}\right.
}
where $k$ is the spring constant ($k\simeq 7.7\times 10^2 {\rm N/m}$) and 
$\Delta_0$ is the pre-compression ($\Delta_0\simeq 1.77\times 10^{-2}{\rm m}$) \footnote{
The values of both $k$ and $\Delta_0$ are directly determined before each spring is confined in a cylinder.
}.
The above form of $f(\delta)$ resembles the power-law form, $f_0 \delta^\alpha,$  with 
$ \alpha \ll 1.$ 

Having fabricated four ``particles''  following the above description,
we prepare {a} bed of plastic straws which are laid transversally to the {axis} of 1D motion.
On this bed the three ``particles'' are aligned in series just in touch with {their} neighbours, see Fig.\ref{fig:data}(b), where the horizontal axis is rotated clockwise by 90 degrees. Towards this target the hitting ``particle'' is manually launched straightforwardly, i.e., along the line of the target cluster. {The Supplementary Materials shows the time evolution (movie1).} 

Fig.\ref{fig:data}(c)  represent the collision experiment of a single particle hitting upon a three-particle  cluster as well as that of a two-particle cluster hitting upon the other two-particle cluster. 
These images has been reconstructed based on the movie taken by a high-speed camera (Pixelink, 2052 frames/s). From each image the horizontal slice containing the thick black marker has been cut out and stuck with the aid of the software ImageJ. The time advances from left to right
and the vertical axis represents the spatial position.
The vertical dotted line in these figures indicates the instant when the hitting particle/cluster are released. The vertical bright-and-dark stripes reflect the period of 10ms thanks to the blinking of fluorescent lamps. The results show well the rigid-body like behavior similar to the numerical model despite the friction by substrate being present. {By analysing the trajectory we estimated the relative deficits in the momentum and energy upon collision to be  5-10\% and 10-20\%,respectively.}

With all the simplicity of the setup that gave the result similar to the numerical ones in the main text,
we have no idea how to extract from (\ref{eq:FvsX}) the corrective coordinate that corresponds to the evanescent mode. Even the very precise value of $\delta$ and $f(\delta)$ near $\delta=0$ are experimentally inaccessible because the insertion of force sensors would alter the mechanical system. We have no {\it a priori} reason to expect an exponential decay of $\delta$'s within the cluster. 
Nevertheless it is certain that the pre-compressed spring is the key ingredient of the rigid-body like behavior:
In fact, if we remove or immobilise the movable piston of each particle, the outcome is a commonly known collision process of ``Newton's cradle'', that is, the furthest particle is dominantly kicked out in exchange of the joining of the hitting particle to the cluster. In the latter case  refined theories and experiments have already been developed \cite{Nesterenko1983, coste-fauve1997, MacKay1999}.

After completing our work we came to notice that the structure of piston and preloaded spring similar to Fig.\ref{fig:data}(a) had been used for the different function:  
For the purpose of attenuating the mechanical shocks between the wagons of the railway trains, they put at the junction of trains  the device called {\it draft gear} \cite{draft-gear-ref1}, which was often combined with frictional elements.
To the authors' knowledge, however, no mention has been made about the aspects of  the collective movement.

\bibliographystyle{apsrev4-1.bst}    \bibliography{ken_LNP_sar}

\end{document}